\documentclass{midl} % Include author names
\usepackage{mwe} % to get dummy images

\jmlrproceedings{MIDL}{Medical Imaging with Deep Learning}
\jmlrpages{}
\jmlryear{2023}

\jmlrworkshop{Short Paper -- MIDL 2023 submission}
\jmlrvolume{-- Under Review}
\editors{Under Review for MIDL 2023}

\title[
Towards Realistic Ultrasound Fetal Brain Imaging Synthesis %Tue  4 Apr 07:28:33 BST 2023
]{
Towards Realistic Ultrasound Fetal Brain Imaging Synthesis %Tue  4 Apr 07:28:33 BST 2023
}

\midlauthor{\Name{Michelle Iskandar\midljointauthortext{Contributed equally}\nametag{$^{2}$}} %\Email{michelle.iskandar@kcl.ac.uk}
\Name{Harvey Mannering\midljointauthortext{Contributed equally}\nametag{$^{1}$}} %\Email{harvey.mannering@ucl.ac.uk}
\Name{Zhanxiang Sun\midljointauthortext{Contributed equally}\nametag{$^{1}$}} %\Email{zhanxiang.sun.22@ucl.ac.uk}
\Name{Jacqueline Matthew \nametag{$^{2}$}} %\Email{xyz@sample.edu}\\
\Name{Hamideh Kerdegari \nametag{$^{2}$}}  % \Email{xyz@sample.edu}\\
\Name{Laura Peralta\nametag{$^{2}$}} %\Email{alphabeta@example.edu}\\
\Name{Miguel Xochicale\midljointauthortext{Contributed equally}\nametag{$^{1}$}} 
\\
\addr $^{1}$ University College London 
\addr $^{2}$ King's College London
}

\begin{document}

\maketitle

\begin{abstract}
Prenatal ultrasound imaging is the first-choice modality to assess fetal health.
Medical image datasets for AI and ML methods must be diverse (i.e. diagnoses, diseases, pathologies, scanners, demographics, etc), however there are few public ultrasound fetal imaging datasets due to insufficient amounts of clinical data, patient privacy, rare occurrence of abnormalities in general practice, and limited experts for data collection and validation.
To address such data scarcity, we proposed generative adversarial networks (GAN)-based models, diffusion-super-resolution-GAN and transformer-based-GAN, to synthesise images of fetal ultrasound brain planes from one public dataset.
We reported that GAN-based methods can generate 256x256 pixel size of fetal ultrasound trans-cerebellum brain image plane with stable training losses, resulting in lower FID values for diffusion-super-resolution-GAN (average 7.04 and lower FID 5.09 at epoch 10) than the FID values of transformer-based-GAN (average 36.02 and lower 28.93 at epoch 60).
The results of this work illustrate the potential of GAN-based methods to synthesise realistic high-resolution ultrasound images, leading to future work with other fetal brain planes, anatomies, devices and the need of a pool of experts to evaluate synthesised images.
Code, data and other resources to reproduce this work are available at \url{https://github.com/budai4medtech/midl2023}.
\end{abstract}


\section{Introduction}
Prenatal imaging is performed to assess various aspects of pregnancy, including confirmation of the pregnancy, screening for developmental defects, and investigation of pregnancy complications~\citep{Kline-Fath2007}.
In the last decade, the fields of machine learning (ML) and artificial intelligence (AI) have been successful to model intelligent behaviors with minimal human interference~\citep{Hamet2017}.
Particularly, automatic classification of fetal ultrasound planes and fetal head biometric measurement~\citep{Burgos-Artizzu2020-1, Sinclair2018, fiorentino2022_arxiv}.
Despite such advances, there are few challenges faced in prenatal imaging:
(a) the accuracy of recorded measurements which can be caused by differences in intra-view variability of imaging equipment and inter-observer variability of sonographer skills~\citep{NHS2015, Sarris2012, Villar1989, Kesmodel2018},
(b) availability of expert clinicians or trained technicians to select, to classify and to validate regions of interest~\citep{Burgos-Artizzu2020},
(c) the insufficient and limited amount of clinical data~\citep{Jang2018, Sinclair2018, He2021}, 
(d) data accessibility due to patient privacy or protection of personal health information~\citep{Shin2018}, and 
(e) the cost of acquisition of clinical data
as it requires expensive imaging equipment and experts for data collection and validation~\citep{Wang2019, Kim2019}.
Given the advances with generative adversarial networks (GAN) methods to handle problems in medical reconstructions, image resolution, enhancement, segmentation, lesion detection, data simulation or classification~\citep{2022_AlAmir_GANMedicalSurvey}, we hypnotise that realistic ultrasound imaging can address challenges in data scarcity, accessibility and expensiveness.
For instance, 
\citet{Hu2017} proposed a method of generating freehand ultrasound image simulation using a spatially conditioned GAN.
\citet{Kazeminia2020} presented a review of the state-of-the-art research in GAN in medical imaging for classification, denoising, reconstruction, synthesis, registration, and detection.
\citet{Montero2021} proposed a method to generate fetal brain US images using an unconditional GAN, StyleGAN2, specifically to improve the fine-grained plane classification, specifically the trans-thalamic and trans-ventricular plane.
% ultrasound fetal imaging~\citep{Korkinof2018_arxiv, Fujioka2019}
Hence, the aim of this work is to show the potential of GAN-based methods to generate realistic ultrasound fetal trans-cerebellum brain plane imaging with small datasets.

\section{Methods and datasets}
\subsection{Diffusion-Super-Resolution-GAN (DSR-GAN)}
We use a Denoising Diffusion Probabilistic Model (DDPM)~\citep{ho2020_neurips} followed by a Super-Resolution-GAN~\citep{Ledig_2017_CVPR}.  To reduce computation time, we finetune a pretrained DDPM to produce 128x128 pixel images and then scale them up to 256x256 using SRGAN.  After the DDPM and before SRGAN, histogram matching \citep{castleman1996digital} is applied to ensure that the colour distribution of the synthetic images matches the colour distribution of the real images.

\subsection{Transformer-based-GAN (TB-GAN)}
We use StyleSwin model, which features Swin Transformer layers designed to capture high quality details of the original images while simultaneously reducing memory usage, enabling synthesized images of higher resolutions ~\cite{Zhang_2022_CVPR}. Differentiable data augmentation (DiffAug) and adaptive pseudo augmentation (APA) are implemented to combat discriminator over-fitting due to limited data and ensure stability in training process ~\cite{zhao2020differentiable, jiang2021deceive}.

\subsection{Image Quality Assessment}
Quality of synthesised images are evaluated with Fr\'echet inception distance (FID), measuring the distance between distributions of synthesised  and original images~\citep{Heusel2017}. 
The lower the FID number is, the more similar the synthesised images are to the original ones. 
FID metric showed to work well for fetal head ultrasound images compared to other metrics ~\cite{Bautista2022}.

\subsection{Datasets}
Trans-cerebellum brain plane ultrasound images from Voluson E6 were used for this work, consisting of 408 training images~\citep{Burgos-Artizzu2020, Burgos-Artizzu2020-1}.
Scans were collected by multiple operators of similar skill level at BCNatal during standard clinical practice between October 2018 and April 2019. 
DICOM images were collected and anominsied using png format, resulting in images of various pixel size (e.g., 692x480, 745x559, and 961x663).

\section{Experiments: Design and results}
\begin{figure}[htbp]
    \centering
    \includegraphics[width=0.99\textwidth]{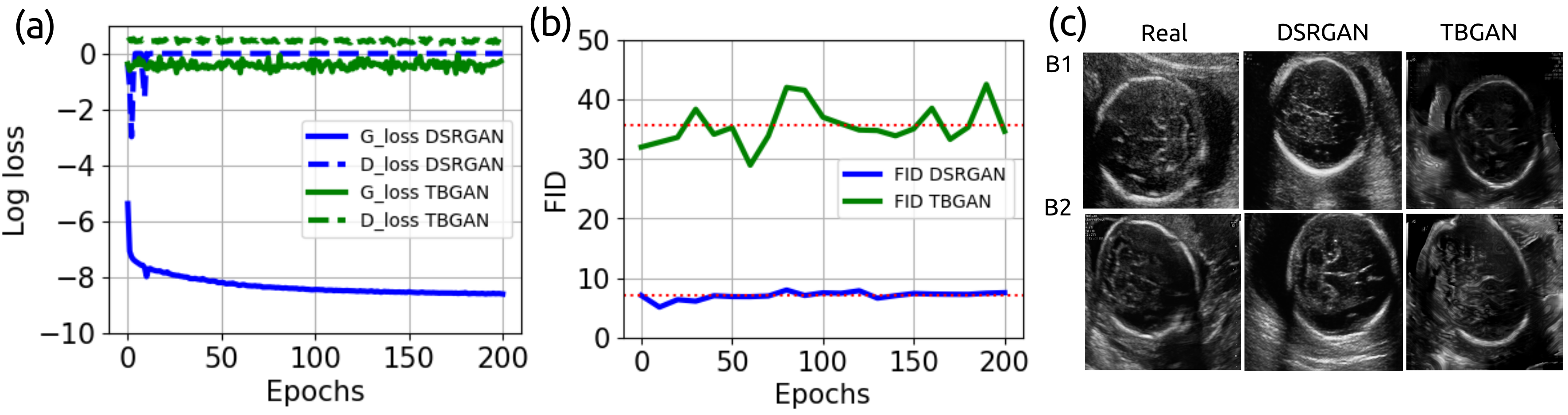} %%ARXIV
\caption{
	Results from Diffusion-Super-resolution-GAN (DSR-GAN) and transformer-based-GAN (TB-GAN):
	(a) Training losses for Generator and Discriminator networks,
	(b) FID scores, and
	(c) 256x256 pixel size trans-cerebellum images of two randomised batches (B1, B2) of real and synthesised (DSR-GAN and TB-GAN).
    }
\label{fig:main_results}
\end{figure}

Diffusion model was finetuned for 10000 epochs with the Adam optimiser to then train SRGAN for 200 epochs from scratch with the Adam optimiser. The images used to train both models are flipped horizontally, zoomed and rotated randomly to increase the variety of the dataset(Fig~\ref{fig:main_results}c).
Transfer learning is used when training StyleSwin. The model was firstly pre-trained on Trans-thalamus plane for 500 epochs as contains larger number of images (1072). Then, the model was fine-tuned on Trans-cerebellum plane images for an additional 200 epochs. Adam optimizer was also used during both pre-training and fine-tuning stages, following the two time-scale update rule with learning rates of 1e-4 for the discriminator and 1e-5 for the generator ~\cite{Heuvel2018}. 

\section{Conclusions and future work}
Synthesising fetal brain images with the diffusion-Super-Resolution-GAN and transformer-based-GAN methods were successful, generating images of 256x256 pixel size resolution with stable loss values and 
resulting in lower FID values for diffusion-super-resolution-GAN (average 7.04 and lower FID 5.09 at epoch 10) compared to FID values of transformer-based-GAN (average 36.02 and lower 28.93 at epoch 60).
Such results suggest, as future work, the potential to synthesise realistic higher-resolution fetal ultrasound images for other anatomies and ultrasound-devices.

% % Acknowledgments---Will not appear in anonymized version
% \midlacknowledgments{
% We thank a bunch of people.
% }

% \newpage 

%%%%%%%%%%%%%%%%%%%%%%%%%%%%%%%%%%%%%
%\bibliography{../references/references}%%%GITHUB
%\bibliography{references}%%%OVERLEAF
%\bibliography{../../references/references}%%%GITHUB/ARXIV

%%%ARXIV

\end{document}